\documentclass[journal=jpccck,layout=onecolumn]{achemso}
\usepackage{multirow}
\usepackage{chemformula} 
\usepackage[T1]{fontenc} 
\usepackage{hyperref}
\usepackage{comment}
\usepackage[version=4]{mhchem}
\usepackage{soul}
\usepackage{lettrine}
\usepackage{siunitx}
\usepackage[version=4]{mhchem}

\author{Djardiel da S. Gomes}
 \affiliation{Universidade Estadual de Campinas, Instituto de F\'{i}sica Gleb Wataghin, Departamento de F\'{i}sica Aplicada, Campinas, S\~ao Paulo, 13083-859, Brazil.}
 \email{djardiel95@gmail.com}
 
 \author{Alexandre F. Fonseca}
 \affiliation{Universidade Estadual de Campinas, Instituto de F\'{i}sica Gleb Wataghin, Departamento de F\'{i}sica Aplicada, Campinas, S\~ao Paulo, 13083-859, Brazil.}
\email{afonseca@ifi.unicamp.br }

\author{Marcelo L. Pereira, Jr}
 \affiliation{University of Bras\'{i}lia, College of Technology, Department of Electrical Engineering, 70910-900, Bras\'{i}lia, Federal District, Brazil.}
 \email{marcelo.lopes@unb.br}

\title[]
  {
  Theoretical Prediction of Three-Dimensional sp$^2$-free Graphyne-Based Nanomaterials via Density Functional Theory
  }


\begin{document}

\begin{abstract}
\noindent The search for carbon-based materials with tailored dimensionality and properties remains an important topic in materials science, particularly for applications in electronics, photonics, and nanomechanics. Among the emerging platforms in this context, graphyne (GY) represents a class of two-dimensional (2D) carbon allotropes composed of benzene rings connected by acetylenic linkages, yielding networks containing both $sp$- and $sp^2$-hybridized carbon atoms. By analogy with the transformation of $sp^2$ carbon networks such as graphene into $sp^3$-bonded diamond through interlayer covalent bonding, we construct three-dimensional (3D) GY-derived frameworks (3DGY) by covalently connecting stacked $\alpha$-, $\beta$-, and $\gamma$-GY sheets via out-of-plane acetylene bridges. This approach converts the original $sp^2$ nodes into $sp^3$ centers while preserving the $sp$ character of the acetylenic segments, producing fully $sp$-$sp^3$ carbon networks. Structural relaxation shows that the $\alpha$-derived framework does not converge to a stable configuration within this scheme, whereas the $\beta$- and $\gamma$-3DGY phases form stable architectures. Density functional theory (DFT) calculations, combined with \textit{ab initio} molecular dynamics (AIMD) simulations, confirm the energetic, thermal, and dynamical stability of these two systems and are further used to investigate their structural, mechanical, electronic, and optical properties. Mechanical analysis reveals anisotropic elastic behavior, whereas electronic structure calculations show indirect band gaps of approximately 0.15~eV for $\beta$-3DGY and 1.65~eV for $\gamma$-3DGY. Optical calculations further reveal anisotropic responses, with absorption extending from the infrared to the visible. These results identify $\beta$-3DGY and $\gamma$-3DGY as new three-dimensional carbon allotropes with distinct mechanical, electronic, and optical properties.
\end{abstract}

\section{Introduction}

Carbon-based materials occupy a central position in materials science. Due to carbon's ability to adopt different hybridizations, namely \textit{sp}, \textit{sp}$^{2}$ and \textit{sp}$^{3}$, a wide variety of stable allotropic structures have been documented \cite{novoselov2004, Hirsch2010}. This structural versatility leads to a remarkable diversity of physical, mechanical, electronic, and optical properties, rendering these materials highly attractive for applications in electronics, photonics, energy storage, and nanomechanics \cite{Geim2009GrapheneReview,Novoselov2012Roadmap,Waris2023EnergyFuels,LOBATOPERALTA2024235140}. Classic allotropes such as diamond \cite{Bundy1996Diamond}, graphite \cite{Dresselhaus2002Graphite}, and more recently, graphene \cite{novoselov2004} exemplify how distinct atomic arrangements of the same element can result in profoundly different physical and chemical properties.

Two-dimensional (2D) materials have attracted enormous attention over the last few decades, largely stimulated by the discovery and rapid development of graphene \cite{Gupta2019ProgMatSci, Terrones2022ACSNanoau, Uddin2023RSCAdv, Paupitz2026Carbon}. Since then, theoretical and computational studies have predicted several other 2D carbon allotropes with distinct bonding topologies, including haeckelites \cite{Terrones2000PRL}, biphenylene networks \cite{Paupitz2012JPCC, Fan2021Science}, phagraphene \cite{wang2015NL}, penta-graphene \cite{zhang2015PNAS}, T-graphene \cite{Liu2012TGraphene}, among others \cite{Ram2018Carbon,Butler2013ACS}. These predictions highlight the remarkable structural diversity that can arise in carbon systems from different bonding arrangements.

Among these families, graphynes (GYs) represent a particularly important class of 2D carbon allotropes. First proposed by Baughman and co-workers \cite{Baughman1987}, GYs can be understood as graphene-like networks in which acetylenic linkages are inserted into the carbon lattice. As a consequence, their structures contain both \textit{sp} and \textit{sp}$^{2}$ hybridized carbon atoms connected through single and triple bonds. This bonding configuration yields more open, porous crystalline frameworks, pronounced structural anisotropy, and greater flexibility for tuning electronic properties compared with graphene \cite{Guerra2024, Nulakani2016}. 

Several structural variants of GYs have been proposed in the literature, among which the $\alpha$-, $\beta$-, and $\gamma$-GY phases are the most widely studied \cite{Baughman1987,Chandran2026GraphyneReview}. These phases differ mainly in the connectivity pattern of the acetylenic segments while preserving the symmetry of the underlying crystalline lattice \cite{Baughman1987, Hou2018}. First-principles calculations, particularly within the framework of density functional theory (DFT), indicate that these materials may display electronic behaviors ranging from metallic to semiconducting, as well as high carrier mobility and tunable optical responses \cite{Chen2013JPCLett, Fazzio2014JPCC, Dani2025MSSP}. Among the predicted phases, the $\gamma$-GY structure has recently been experimentally synthesized \cite{Valentin2022JACS,barua2022Carbon}. These characteristics reinforce the potential of GYs as versatile materials for nanoelectronic and optoelectronic applications \cite{Puigdollers2016, Soni2014, Chen2020, Sani2020, Azizi2023}.

Despite the remarkable properties of 2D materials, their reduced dimensionality may limit certain practical applications. For example, some 2D metals may become more stable than their 3D counterparts only at very small length scales \cite{Wang2020MatTodAdv}. In addition, emergent phenomena associated with Moir\'{e} patterns in bilayer and trilayer systems cannot be reproduced in isolated monolayers \cite{Shreyas2023CPI, Ren2025JPhysCond}. More generally, dimensionality has been recognized as a key parameter for tailoring the properties of carbonaceous materials \cite{Li2019AppEne, Xiao2021AdvEneMat, Wang2025CSR}. In this context, the construction of three-dimensional (3D) architectures from 2D systems has emerged as a promising strategy to expand the range of accessible properties and technological applications. \cite{Hasani2025, Wang2016, Gao2021, Ling2013, Menghao2011}.

Recent theoretical studies have explored various routes to generate 3D carbon structures from 2D precursors. For instance, TH-graphyne has been proposed as a new porous 2D carbon allotrope \cite{lima2025th}, and subsequent investigations have examined its corresponding 3D form as a promising porous carbon framework \cite{hassan2025unveiling}. Similarly, Irida-graphene was introduced as a 2D carbon allotrope \cite{junior2023irida}, and later extended to 3D nanostructures through interlayer covalent connections \cite{felix20253d}. Other theoretical approaches have also demonstrated that 2D carbon allotropes, such as biphenylene-based structures, can be transformed into 3D networks through topological mapping strategies \cite{tromer2024transforming}. Together, these studies illustrate the growing interest in designing three-dimensional carbon materials derived from low-dimensional precursors.

One possible route to achieve this transformation is to assemble layered structures through stacking or substrate-supported architectures. An alternative strategy is to design direct chemical connections between the layers, thereby generating fully covalent three-dimensional frameworks. In carbon systems, examples of such approaches include polyyne-based diamond frameworks \cite{Douglas1993Nature} and families of the so-called $n$-diamondynes \cite{costa2018Carbon}. The synthesis of one form of diamondyne has recently been reported \cite{Yang2025Chemrxiv}, and computational studies have explored the structural and physical properties of broader diamondyne families \cite{Bastos2025ACSOmega}. These developments demonstrate that acetylene-based bonding motifs provide a viable pathway for constructing three-dimensional carbon networks derived from low-dimensional precursors.

In this context, GY networks constitute particularly attractive building blocks for the design of new three-dimensional carbon allotropes. Their intrinsic combination of \textit{sp} and \textit{sp}$^{2}$ hybridizations offers natural sites for structural reconfiguration through the formation of additional covalent bonds. 
Indeed, there has been considerable research on all-\textit{sp}$^{2}$ and \textit{sp}$^{2}+$\textit{sp}$^{3}$ 3D carbon phases that can be derived from connecting layers of graphyne.~\cite{Belenkov2011Russian,Hu2014JSuperhard,Bu2014JMCC,Yu2018AIPAdv,Ma2023PCCP,FonsecaRay2025PNAS}
These studies revealed noteworthy properties, including metallic behavior and superhardness. 
To the best of our knowledge, aside from the diamondyne-like structures~\cite{Douglas1993Nature,costa2018Carbon,Yang2025Chemrxiv,Bastos2025ACSOmega}, there is no other design of extended 3D frameworks from planar networks that preserves the acetylenic motifs characteristic of GY structures.

In this work, 3D allotropes derived from GY (3DGY) phases are introduced. These structures are obtained by covalently connecting stacked GY layers through out-of-plane acetylene bridges. This construction converts the original \textit{sp}$^{2}$ nodes of the parent GY sheets into \textit{sp}$^{3}$ centers while preserving the \textit{sp} character of the acetylenic segments, resulting in fully \textit{sp}-\textit{sp}$^{3}$ carbon networks. In this sense, these architectures can be regarded as conceptual analogues of the transformation that relates graphene and diamond, but originating from GY frameworks. The resulting structures form continuous 3D covalent networks without weakly coupled planes and remain entirely free of \textit{sp}$^{2}$ hybridization. This structural motif defines a new class of carbon allotropes characterized by relatively open frameworks and reduced density, which enables the emergence of distinct mechanical, electronic, and optical properties and expands the landscape of three-dimensional carbon materials beyond the classical diamond paradigm.

\section{Methodology}

The 3D GY-derived structures investigated in this work were constructed from stacked layers of GY phases. In particular, the $\alpha$-, $\beta$-, and $\gamma$-GY lattices were initially considered structural precursors. The construction strategy consists of establishing covalent connections between adjacent GY layers through out-of-plane acetylenic bridges, thereby transforming the original planar networks into fully 3D frameworks. 

In the parent GY structures, specific carbon atoms exhibit $sp^{2}$ hybridization and act as junction points connecting the acetylenic segments. In the present construction scheme, these nodes serve as anchoring sites for interlayer bonding. When an additional acetylenic unit is attached perpendicular to the original layer, the coordination of these carbon atoms increases from three to four, converting their hybridization from $sp^{2}$ to $sp^{3}$. At the same time, the linear geometry of the acetylenic segments preserves the $sp$ character of the triple-bonded carbon atoms. As a result, the resulting 3D frameworks become entirely free of $sp^{2}$ hybridization and consist exclusively of $sp$ and $sp^{3}$ carbon atoms.

The specific geometrical arrangement of these interlayer connections depends on the topology of the parent GY lattice. For the $\alpha$-GY phase, each $sp^{2}$ node connects three acetylenic pairs within the plane. In the proposed construction, additional acetylenic bridges were introduced in an alternating fashion, so that each node connects to either the upper or the lower layer. This configuration converts the original threefold-coordinated carbon atoms into fourfold-coordinated centers, thereby producing the intended $sp^{3}$ hybridization while preserving the linear $sp$ character of the acetylenic chains. However, structural relaxation calculations indicate that the resulting $\alpha$-derived 3D framework does not preserve the intended topology. During the optimization process, significant lattice distortions arise, leading to the loss of the original structural connectivity and preventing convergence to a stable configuration within this bonding scheme.

For the $\beta$-GY phase, pairs of neighboring $sp^{2}$ nodes are present in the structure. In this case, acetylenic bridges are attached alternately to these adjacent atoms, with one bridge pointing toward the upper layer and the other toward the lower layer. This arrangement promotes the formation of stable tetrahedral coordination around the transformed $sp^{3}$ centers while maintaining structural symmetry along the stacking direction.

For the $\gamma$-GY phase, the structure contains six $sp^{2}$ carbon atoms arranged in benzene-like rings. In this configuration, the interlayer bridges are distributed in an alternating manner around the ring, with three acetylenic pairs connecting to the upper layer and the remaining three connecting to the lower layer. This symmetric distribution ensures that the numbers of bonds pointing upward and downward are equal, thereby preserving the overall structural balance of the three-dimensional lattice.

\begin{figure}[t!]
    \centering
    \includegraphics[width=1.0\linewidth]{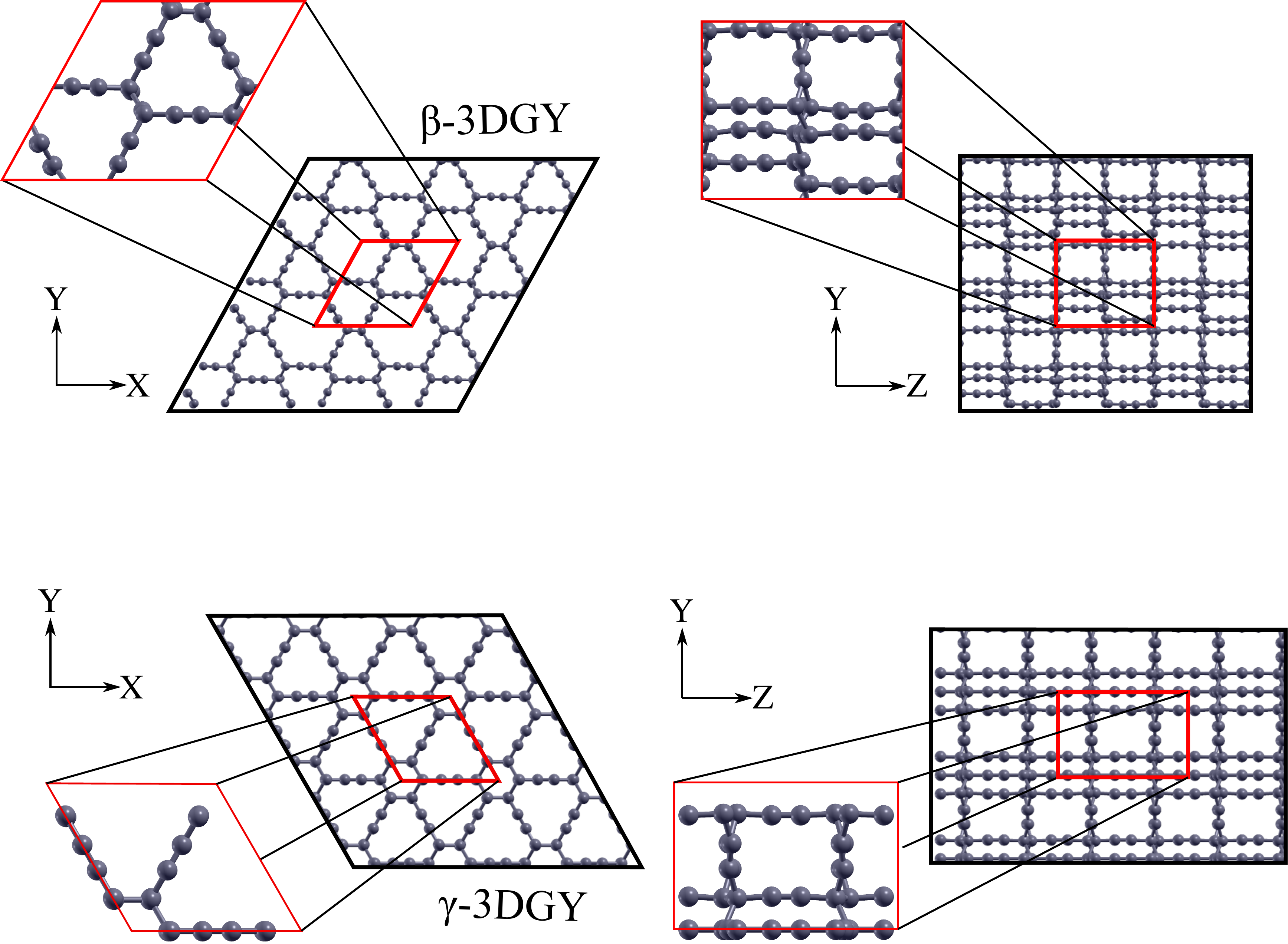}
    \caption{Optimized structures of the $\beta$-3DGY and $\gamma$-3DGY phases obtained from stacked graphyne layers connected through out-of-plane acetylenic bridges. The red lines highlight the corresponding unit cells.}
    \label{fig01}
\end{figure}

The fully optimized structures obtained after geometry relaxation are shown in Figure~\ref{fig01}. The figure illustrates the atomic arrangements of the stable $\beta$-3DGY and $\gamma$-3DGY phases, together with the orientation of the acetylenic bridges that connect the layers. The red lines highlight the corresponding unit cells used in the simulations. The $\alpha$-derived structure is not shown because the optimization procedure substantially reconstructs the lattice, thereby preventing the preservation of the proposed three-dimensional topology. For this reason, the $\alpha$-based configuration was not considered in the subsequent analyses, and only the $\beta$-3DGY and $\gamma$-3DGY structures were investigated throughout this work. The crystallographic information files for the stable structures are available in the supporting information.

The study of geometry optimizations, structural stability, and electronic and optical properties of the $\beta$-3DGY and $\gamma$-3DGY structures was carried out using first-principles calculations implemented in the SIESTA code \cite{Soler2002, Garcia2020}, within the framework of density functional theory (DFT) \cite{Kohn1965, Sanchez1997}. The exchange-correlation effects were described using the generalized gradient approximation (GGA) with the Perdew-Burke-Ernzerhof (PBE) functional \cite{Perdew1996}. Electron-ion interactions were treated using norm-conserving Troullier-Martins pseudopotentials in the Kleinman-Bylander factorized form \cite{Troullier1991, Kleinman1982}, considering the valence electronic configuration $2s^{2}2p^{2}$ for carbon atoms.

For the geometric optimizations, a mesh cutoff energy of 800 Ry and a double-$\zeta$ polarized (DZP) basis set composed of finite-range numerical atomic orbitals were employed. Brillouin zone integrations were performed using a $10 \times 10 \times 10$ Monkhorst-Pack $k$-point grid \cite{Monkhorst1976}. During the structural relaxation, both the lattice vectors and atomic positions were fully optimized until the maximum force on each atom was smaller than $10^{-3}$ eV/\AA\ and the total energy variation between consecutive steps was below $10^{-5}$ eV.

The dynamical stability of the optimized structures was evaluated through phonon calculations and \textit{ab initio} molecular dynamics (AIMD) simulations. The unit cells contain 52 and 36 atoms for the $\beta$-3DGY and $\gamma$-3DGY structures, respectively. Phonon calculations were performed using interpolation across the Brillouin zone with a mesh cutoff of 700 Ry. Convergence criteria of $10^{-5}$ eV for energy and $0.001$ eV/\AA\ for forces were adopted. The acoustic sum rule was also applied in order to ensure correct vibrational behavior at the $\Gamma$ point.

Thermal stability was further investigated through AIMD simulations using a $2 \times 2 \times 2$ supercell. Simulations were performed in the canonical ensemble (NVT) using the Nosé thermostat \cite{tromer2020} with a time step of 1 fs. The systems were simulated at temperatures of 300 K, 600 K, 900 K, and 1000 K. Each simulation was gradually heated to the target temperature and then evolved for a total simulation time of 7 ps to monitor possible structural distortions or bond rearrangements.

The elastic properties of the structures were obtained by applying incremental uniaxial strains of $0.5\%$ until structural fracture. Young's modulus was calculated from the slope of the linear region of the stress-strain curve, typically between $0$ and $1\%$ strain. The Poisson ratio was determined from the ratio between the transverse strain $\varepsilon_{\text{trans}}$ and the axial strain $\varepsilon_{\text{axial}}$ for small deformations, following the relation proposed by Kang \cite{Kang2011}:
\begin{equation}
\nu = -\frac{\varepsilon_{\text{trans}}}{\varepsilon_{\text{axial}}}.
\end{equation}
The optical properties were evaluated by applying an external electric field of $1.0$ V/\AA\ independently along the $x$, $y$, and $z$ directions. Using the Kramers-Kronig relations together with Fermi's golden rule, the real $(\epsilon_1)$ and imaginary $(\epsilon_2)$ parts of the dielectric function were calculated. The real part is given by
\begin{equation}
\epsilon_1(\omega)=1+\frac{1}{\pi}P\int_{0}^{\infty}d\omega'\frac{\omega'\epsilon_2(\omega')}{\omega'^{2}-\omega^{2}},
\end{equation}
\noindent where $P$ denotes the principal value of the integral. The imaginary part, which accounts for interband optical transitions between the valence band (VB) and the conduction band (CB), is expressed as
\begin{equation}
\begin{split}
\epsilon_2(\omega)&=\frac{4\pi^2}{\Omega\omega^2} \\
\times&
\sum_{\substack{i\in \text{VB}\\ j\in \text{CB}}}
\sum_{k}W_k |\rho_{ij}|^2 
\delta(\epsilon_{kj}-\epsilon_{ki}-\hbar\omega).
\end{split}
\end{equation}

From the calculated values of $\epsilon_1$ and $\epsilon_2$, additional optical quantities were obtained, including the absorption coefficient $\alpha$, the reflectance $R$, and the refractive index $\eta$:
\begin{equation}
\alpha(\omega)=\sqrt{2}\omega\left [ (\epsilon_1^{2}(\omega)+\epsilon_2^{2}(\omega))^{1/2}-\epsilon_1(\omega)\right ]^{1/2},
\end{equation}
\begin{equation}
R(\omega)=\left [ \frac{(\epsilon_1(\omega)+i\epsilon_2(\omega))^{1/2}-1}{(\epsilon_1(\omega)+i\epsilon_2(\omega))^{1/2}+1} \right ]^2,
\end{equation}
and
\begin{equation}
\eta(\omega)=\frac{1}{\sqrt{2}}\left [ (\epsilon_1^{2}(\omega)+\epsilon_2^{2}(\omega))^{1/2}+\epsilon_1(\omega) \right ]^{2}.
\end{equation}

\section{Results and Discussion}

In this section, we present and discuss the structural, mechanical, electronic, and optical properties of the three-dimensional graphyne-derived frameworks identified as stable in our construction scheme. After structural relaxation and stability verification, the analysis focuses on how the topology of the $\beta$-3DGY and $\gamma$-3DGY networks influences their mechanical response, electronic structure, and optical behavior. These results provide a comprehensive characterization of the physical properties emerging from the $sp$–$sp^3$ carbon bonding configuration in these three-dimensional architectures.

\subsection{Structural properties and stability}

The $\beta$-3DGY and $\gamma$-3DGY structures discussed here correspond to the optimized configurations obtained from the construction strategy described in the Methodology section and illustrated in Figure~\ref{fig01}. The unit cell of the $\beta$-3DGY structure contains 56 carbon atoms, whereas the $\gamma$-3DGY unit cell contains 36 atoms.

The optimized lattice parameters indicate distinct crystallographic symmetries for the two structures. The $\beta$-3DGY phase presents lattice parameters $a=b=c=9.63$~\AA\ with angles $\alpha=\beta=90^{\circ}$ and $\gamma=61^{\circ}$, suggesting a distorted rhombohedral-like geometry. In contrast, the $\gamma$-3DGY structure exhibits lattice parameters $a=b=7.16$~\AA\ and $c=9.44$~\AA\ with angles $\alpha=\beta=90^{\circ}$ and $\gamma=120^{\circ}$, consistent with a hexagonal symmetry inherited from the parent $\gamma$-GY lattice.

Both structures preserve the characteristic acetylenic linkages of graphyne networks. The $-\ce{C#C}-$ bonds exhibit typical triple-bond lengths of approximately 1.20~\AA, while the adjacent $\sigma$ bonds connecting the acetylenic units present lengths around 1.36~\AA. These values are consistent with bond lengths reported for other graphyne-derived systems and indicate that the linear geometry of the $sp$-hybridized carbon atoms is preserved after the formation of the three-dimensional frameworks.

The previously described construction strategy converts the $sp^{2}$- hybridized junction atoms of the parent graphyne sheets into $sp^{3}$ coordination centers by forming interlayer covalent bonds. Consequently, the resulting 3D structures become fully composed of $sp$ and $sp^{3}$ carbon atoms, without the presence of $sp^{2}$ hybridization. This structural transformation yields relatively open three-dimensional frameworks compared with dense carbon allotropes such as diamond, while retaining the acetylenic motifs characteristic of graphyne networks.

\begin{figure}[t!]
\centering
\includegraphics[width=1.0\linewidth]{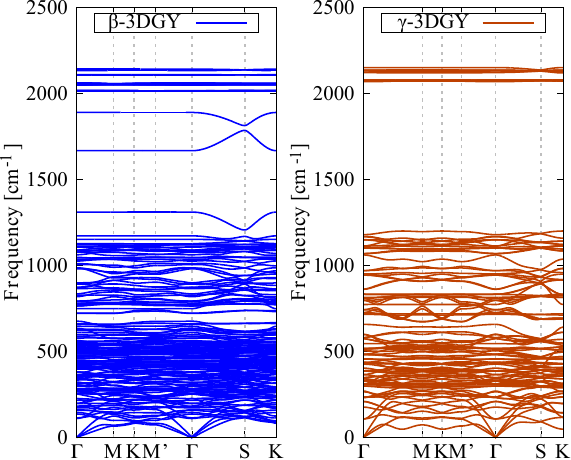}
\caption{Phonon dispersion curves for $\beta$-3DGY and $\gamma$-3DGY structures.}
\label{fig02}
\end{figure}

To evaluate the dynamical stability of these structures, phonon dispersion calculations were performed. The phonon spectra obtained for the $\beta$-3DGY and $\gamma$-3DGY phases are presented in Figure~\ref{fig02}. These calculations were carried out using $2\times2\times2$ supercells containing 416 and 288 atoms for the $\beta$-3DGY and $\gamma$-3DGY structures, respectively.

Inspection of Figure~\ref{fig02} reveals the absence of imaginary frequencies throughout the entire Brillouin zone for both structures. This result indicates that the optimized geometries correspond to dynamically stable configurations. Additionally, vibrational modes above 2000~cm$^{-1}$ are observed, which are characteristic of the stretching vibrations of the acetylenic $-\ce{C#C}-$ bonds. The presence of these high-frequency modes confirms the preservation of the triple-bonded carbon segments in the three-dimensional structures.

In addition to dynamical stability, the thermal stability of the structures was investigated through \textit{ab initio} molecular dynamics (AIMD) simulations. The time evolution of the total energy per atom and the temperature fluctuations were monitored during simulations performed in the NVT ensemble. Supercells of size $2\times2\times2$ were employed, and the systems were simulated for a total time of 7 ps.

\begin{figure*}[t]
\centering
\includegraphics[width=0.9\linewidth]{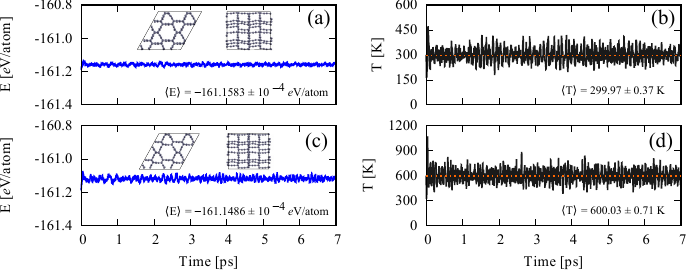}
\caption{Time evolution of the total energy per atom for $\beta$-3DGY lattices at 300 K and 600 K during 7 ps of AIMD simulation. The inset figures show the top and side views of the final structures.}
\label{fig03}
\end{figure*}

\begin{figure*}[t]
\centering
\includegraphics[width=0.9\linewidth]{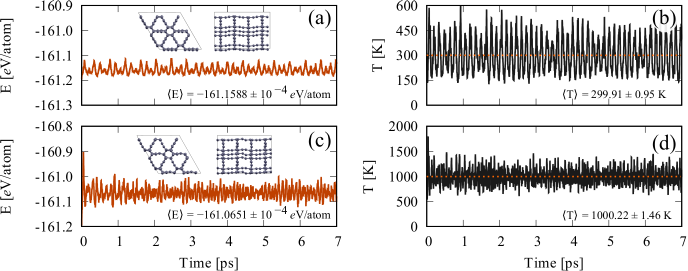}
\caption{Time evolution of the total energy per atom for $\gamma$-3DGY lattices at 300 K and 1000 K during 7 ps of AIMD simulation. The inset figures show the final atomic configurations.}
\label{fig04}
\end{figure*}

For the $\beta$-3DGY structure, simulations were performed at temperatures of 300 K and 600 K. The corresponding results are shown in Figure~\ref{fig03}. The total energy per atom oscillates about a nearly constant average over the entire simulation, while temperature fluctuations remain centered around the target values. The final atomic configurations shown in the inset images indicate that the structural topology is preserved, with no evidence of bond breaking or structural reconstruction.

A similar analysis was performed for the $\gamma$-3DGY phase, with simulations conducted at temperatures from 300 K to 1000 K. The results are presented in Figure~\ref{fig04}. Even at temperatures as high as 1000 K, the energy fluctuations remain small, and the temperature remains stable throughout the simulation.

The final snapshots obtained from the AIMD simulations confirm that the three-dimensional framework of the $\gamma$-3DGY structure remains intact even at elevated temperatures. No bond breaking or structural collapse is observed during the simulations. These results demonstrate that both $\beta$-3DGY and $\gamma$-3DGY phases exhibit not only dynamical stability, as confirmed by phonon calculations, but also significant thermal stability under the investigated temperature conditions.

\subsection{Elastic Properties}

To investigate the mechanical response of the 3DGY structures, uniaxial tensile deformations were applied along the crystallographic $x$, $y$, and $z$ directions in strain increments of $0.1\%$.

\begin{figure*}
    \centering
    \includegraphics[width=0.7\linewidth]{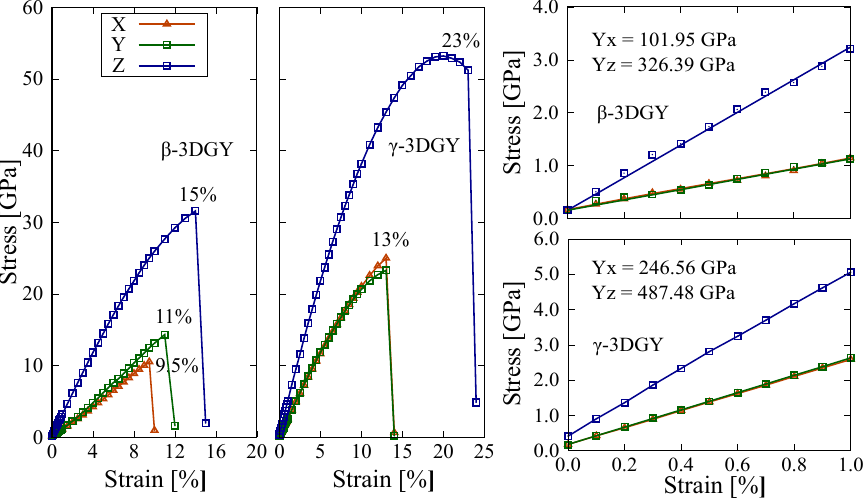}
    \caption{Stress-strain curves of $\beta$-3DGY and $\gamma$-3DGY under uniaxial deformation along the $x$, $y$, and $z$ directions. The left panels show the full stress-strain response up to fracture, while the right panels present enlarged views of the elastic regime ($0$-$1\%$ strain) used to extract the Young's modulus.}
    \label{fig041}
\end{figure*}

The resulting stress-strain curves are presented in Figure~\ref{fig041}. The left panels display the full mechanical response of $\beta$-3DGY and $\gamma$-3DGY up to fracture for the three loading directions, whereas the right panels show magnified views of the elastic regime ($0$-$1\%$ strain), from which the Young's moduli were obtained.

As observed in Figure~\ref{fig041}, both structures exhibit identical responses along the $x$ and $y$ directions, indicating isotropic elastic behavior within the $xy$ plane. This symmetry originates from the topology of the parent graphyne layers, which preserves equivalent bonding environments along these two directions even after the formation of the three-dimensional network.

The Young's modulus values were extracted from the linear region of the curves shown in the right panels of Figure~\ref{fig041}. For $\beta$-3DGY, the in-plane stiffness is $Y_x = Y_y = 101.95$ GPa, while the out-of-plane modulus reaches $Y_z = 326.39$ GPa. In contrast, $\gamma$-3DGY presents significantly larger elastic constants, with $Y_x = Y_y = 246.56$ GPa and $Y_z = 487.48$ GPa. These results indicate that the $\gamma$ phase is mechanically stiffer than the $\beta$ phase in all crystallographic directions.

A pronounced elastic anisotropy is also evident from the stress-strain curves. In both materials the stress increases more rapidly when the deformation is applied along the $z$ direction, reflecting the higher rigidity associated with the vertical covalent connections formed by the acetylenic bridges. The ratios $Y_z/Y_x$ are approximately $3.2$ for $\beta$-3DGY and $2.0$ for $\gamma$-3DGY, confirming that both frameworks are considerably stiffer along the stacking direction than within the plane.

The Poisson ratios further highlight differences between the two structures. In the $xy$ plane the Poisson ratio values are $\nu_{yx} = 0.17$ for $\beta$-3DGY and $\nu_{yx} = 0.01$ for $\gamma$-3DGY. The coupling between in-plane and out-of-plane deformation is given by $\nu_{xz} = 0.13$ and $0.05$ for $\beta$- and $\gamma$-3DGY, respectively.

The fracture behavior, also visible in Figure~\ref{fig041}, reveals distinct mechanical resilience between the two structures. For $\beta$-3DGY, failure occurs at strains of approximately $9.5\%$ and $11.0\%$ along the $x$ and $y$ directions, respectively, while the maximum strain along the $z$ direction reaches about $15\%$. In contrast, $\gamma$-3DGY sustains larger tensile deformation, fracturing at about $13\%$ in the plane and up to $23\%$ along the $z$ axis. This indicates a significantly greater mechanical robustness of the $\gamma$ structure, particularly for loading perpendicular to the layers.

Inspection of the fracture patterns indicates that the initial structural failure in both materials originates from the rupture of $\sigma$ bonds connecting the acetylenic segments. These bonds constitute the mechanically weakest elements of the three-dimensional network and their rupture triggers the abrupt stress drops observed in the curves.

\subsection{Electronic Behavior}

\begin{figure}[t!]
    \centering
    \includegraphics[width=1.0\linewidth]{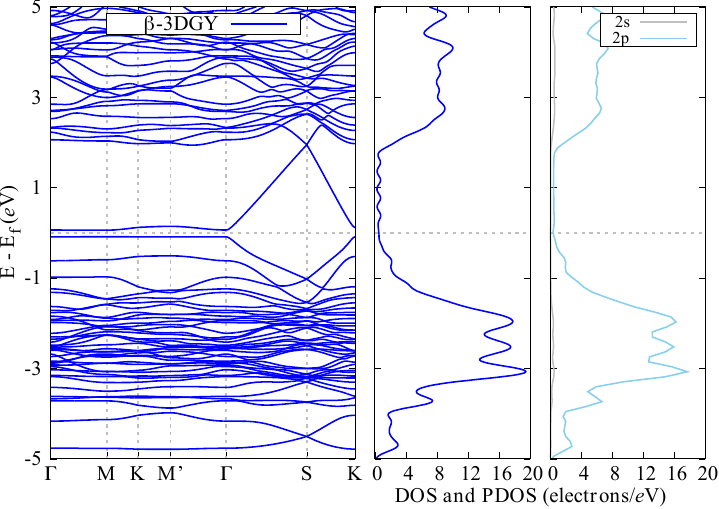}
    \caption{Electronic band structure (left), total density of states (middle), and projected density of states (right) of $\beta$-3DGY calculated using the PBE functional.}
    \label{fig05}
\end{figure}

\begin{figure}[t!]
    \centering
    \includegraphics[width=1.0\linewidth]{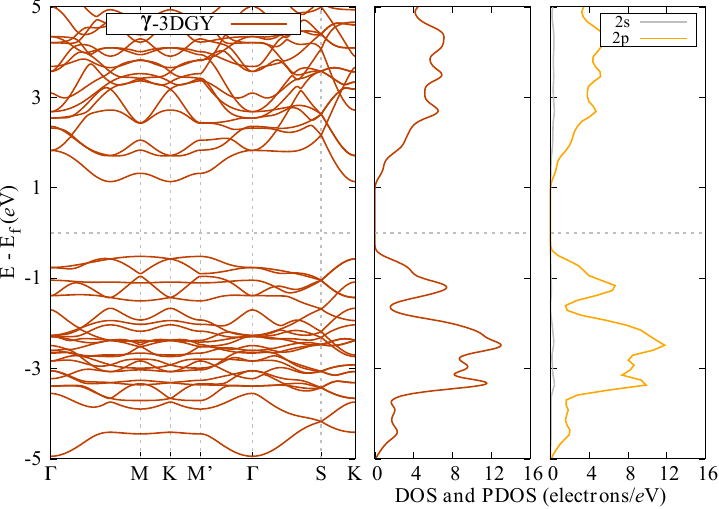}
    \caption{Electronic band structure (left), total density of states (middle), and projected density of states (right) of $\gamma$-3DGY calculated using the PBE functional.}
    \label{fig06}
\end{figure}

The electronic properties of the proposed structures were investigated using band-structure calculations and total and projected density-of-states analyses. The results for $\beta$-3DGY and $\gamma$-3DGY are shown in Figures~\ref{fig05} and \ref{fig06}, respectively. In each figure, the left panel presents the electronic band structure along the high-symmetry path of the Brillouin zone, the central panel shows the total density of states (DOS), and the right panel displays the projected density of states (PDOS) resolved into the carbon $2s$ and $2p$ orbitals.

As shown in Figure~\ref{fig05}, the $\beta$-3DGY structure exhibits semiconducting behavior with an indirect band gap of approximately $0.15$ eV. The valence band maximum and conduction band minimum occur at different $k$-points in the Brillouin zone, confirming the indirect nature of the gap. The band dispersion near the Fermi level is relatively moderate, which is consistent with the presence of extended $\pi$-type interactions along the carbon network. 

The DOS profile further confirms the narrow-gap semiconducting character of the structure. In particular, the electronic states near the Fermi level have a relatively low density, indicating that only a small number of states contribute to charge transport in this energy range. This feature suggests that $\beta$-3DGY behaves as a narrow-gap semiconductor with a limited density of intrinsic charge carriers.

The PDOS analysis reveals that the electronic states near the valence and conduction band edges are dominated by the carbon $2p$ orbitals, whereas the contribution from the $2s$ orbitals is significantly smaller in this energy range. This behavior reflects the dominant role of $p$-type bonding in the electronic structure, which originates from the presence of acetylenic segments and conjugated carbon chains within the framework. The strong $2p$ contribution is therefore consistent with the mixed $sp$-$sp^{3}$ hybridization present in the three-dimensional graphyne-derived network.

The electronic structure of $\gamma$-3DGY, shown in Figure~\ref{fig06}, also displays semiconducting behavior but with a significantly larger indirect band gap of approximately $1.65$ eV. Compared to $\beta$-3DGY, the conduction and valence bands are more widely separated, indicating a stronger electronic localization within the lattice.

The DOS distribution corroborates this behavior by showing a clear energy window devoid of electronic states near the Fermi level, consistent with a typical semiconductor with a moderate band gap. The increased gap width suggests that $\gamma$-3DGY may present improved electronic stability against thermally activated carriers when compared to the $\beta$ phase.

Similar to the $\beta$ structure, the PDOS analysis indicates that the electronic states near the band edges are primarily derived from the carbon $2p$ orbitals. This result highlights that the electronic structure of both materials is governed by the $p$-orbital network associated with the carbon framework, while the $2s$ orbitals mainly contribute to deeper valence states.

Overall, the comparison between the two systems indicates that the topology of the three-dimensional graphyne-derived framework strongly influences the electronic behavior. While $\beta$-3DGY behaves as a narrow-gap semiconductor with a very small band gap, $\gamma$-3DGY exhibits a significantly larger band gap and therefore a more pronounced semiconducting character. These results suggest that structural variations within the 3D graphyne family can be used to tune the electronic properties of carbon-based porous frameworks.

\subsection{Optical Response}

\begin{figure*}[t!]
    \centering
    \includegraphics[width=0.7\linewidth]{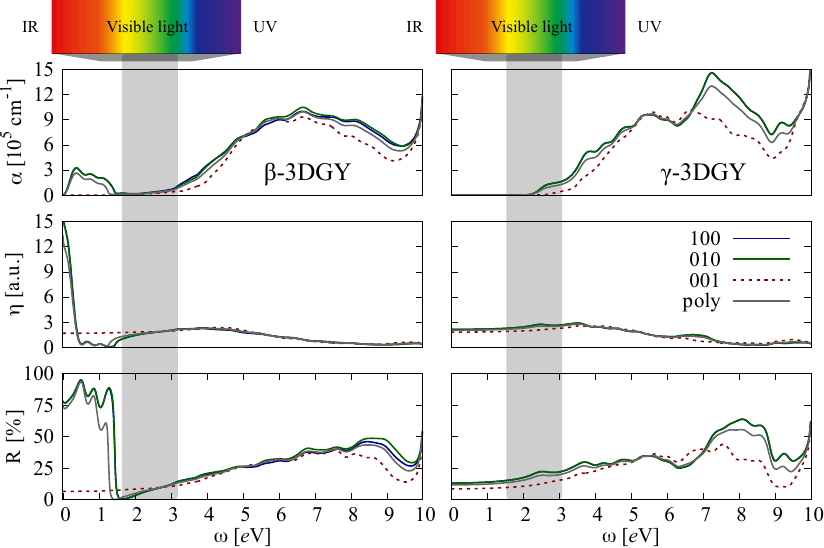}
    \caption{Optical properties of $\beta$-3DGY and $\gamma$-3DGY calculated using the PBE functional. The panels show the absorption coefficient ($\alpha$), refractive index ($\eta$), and reflectivity ($R$) as functions of photon energy for different light polarizations. The shaded region highlights the visible spectrum.}
    \label{fig07}
\end{figure*}

The optical response of the $\beta$-3DGY and $\gamma$-3DGY structures was investigated through the calculation of the absorption coefficient ($\alpha$), the refractive index ($\eta$), and the reflectivity ($R$) as functions of photon energy. The results are presented in Figure~\ref{fig07}, where the optical spectra are shown for light polarized along the crystallographic directions [100], [010], and [001], together with the polycrystalline average. All optical properties were obtained within the GGA/PBE approximation.

The top panels of Figure~\ref{fig07} display the absorption coefficient spectra for both structures. In general, the optical activity of the two materials is concentrated mainly in the ultraviolet region, while only a small portion of the absorption occurs within the visible range. This behavior is consistent with the semiconducting character obtained from the electronic band structures. Using the Tauc plot method, the optical band gaps were estimated as $(0.2100 \pm 0.005)$ eV for $\beta$-3DGY and $(1.6200 \pm 0.034)$ eV for $\gamma$-3DGY, values that closely match the electronic band gaps obtained from the band structure calculations.

The spectra also reveal that the optical response is essentially identical along the [100] and [010] directions, indicating in-plane optical isotropy. This result reflects the symmetry of the crystal lattice within the $xy$ plane, which produces equivalent electronic transitions for light polarized along these directions. In contrast, the response along the [001] direction is significantly weaker, indicating a pronounced optical anisotropy between the in-plane and out-of-plane directions. This behavior is directly related to the structural anisotropy introduced by the acetylenic bridges that connect the stacked graphyne layers.

The middle panels of Figure~\ref{fig07} present the refractive index spectra. In the visible range, the refractive index remains nearly constant for both materials, indicating that no significant electronic transitions occur in this energy interval. The onset of noticeable variations in $\eta$ occurs only above approximately $3$ eV, where interband electronic transitions begin to contribute significantly to the optical response. This result is consistent with the absorption spectra, which show stronger optical activity in the ultraviolet region.

The bottom panels show the reflectivity spectra for the two materials. The $\beta$-3DGY structure exhibits relatively high reflectivity in the infrared region, followed by a sharp decrease as the photon energy approaches the visible range. This behavior indicates that the material becomes increasingly transparent as the photon energy increases toward the visible spectrum. In contrast, the $\gamma$-3DGY structure displays a gradual increase in reflectivity with increasing photon energy, particularly in the ultraviolet region.

Indeed, both structures exhibit low reflectivity and weak absorption in the visible range, indicating that they are largely transparent in this spectral region. This combination of optical transparency in the visible region and strong absorption in the ultraviolet range suggests that three-dimensional graphyne-derived frameworks may present potential for applications in optoelectronic devices operating in the UV region.

\section{Conclusions}

In this work, two three-dimensional graphyne-derived carbon allotropes, $\beta$-3DGY and $\gamma$-3DGY, were investigated using first-principles calculations. Both structures were found to be dynamically stable and exhibit pronounced anisotropic mechanical behavior, with higher stiffness along the out-of-plane direction due to the acetylenic interlayer connections. Notably, $\gamma$-3DGY presents a near-zero Poisson's ratio, a relatively rare mechanical feature that may enable potential applications in aeronautical and biomedical materials \cite{fortes1989MSEA,fonseca2020PSSRRL, Chen2015AST, Freed2008NatMat, Soman2012SoftMat}.

Electronic structure calculations reveal that both systems behave as indirect semiconductors, with a narrow band gap for $\beta$-3DGY and a significantly larger gap for $\gamma$-3DGY. The optical response shows low reflectivity and weak absorption in the visible region, with stronger activity in the ultraviolet range. These results indicate that structural variations within three-dimensional graphyne frameworks provide an effective route to tune the mechanical, electronic, and optical properties of carbon-based materials.

\begin{acknowledgement}
This study was financed in part by the Coordination for the Improvement of Higher Education Personnel (CAPES), the National Council for Scientific and Technological Development (CNPq), the São Paulo Research Foundation (FAPESP), and the Research Support Foundation of the Federal District (FAPDF). D.G.S. acknowledges financial support from CAPES under grant No. 88887.102348/2025-00 (Finance Code 001). A.F.F. is a fellow of CNPq under grant No. 302009/2025-6 and also acknowledges support from FAPESP under grant No. 2024/14403-4. M.L.P.J. acknowledges financial support from FAPDF under grant No. 00193-00001807/2023-16, from CNPq under grants No. 444921/2024-9 and 308222/2025-3, and from CAPES under grant No. 88887.005164/2024-00.
\end{acknowledgement}

\begin{suppinfo}
The Supporting Information contains the crystallographic information files (CIF) of the optimized $\beta$-3DGY and $\gamma$-3DGY structures.
\end{suppinfo}

\bibliography{references}

\end{document}